# Including $\beta-$ and $\gamma-$ Vibrations in Evaluating the Ground State Rotational Band of Deformed Even-Even Nuclei

Mohamed E. Kelabi[*]
A. Y. Ahmed[*]
Vikram Singh[†]

**Abstract**
From the semi-empirical formalisms of Bohr-Mottelson, a new model, based on the effect of $\beta$- and $\gamma$- head energies and the variable moment of inertia, was developed to calculate the ground state rotational band of almost all deformed *e-e* nuclei. The model can be tuned for nuclei with experimentally available $\beta$- and $\gamma$-vibrational energies as well as for those with unmeasured $\beta$- and $\gamma$-vibrations.

**Introduction**
The Rotational-Vibrational model RVM of Bohr-Mottelson[1]

$$E(J) = AJ(J+1) - BJ^2(J+1)^2 \qquad (1)$$

has been modified to take the effect of nuclear rotations via the variable moment of inertia[2], and introducing a correction term due to $\beta$- and $\gamma$- vibrations[3]

$$\hbar\omega_\beta = \left[\frac{\tfrac{3}{2}A^3}{B - \tfrac{1}{2}A^3/(\hbar\omega_\gamma)^2}\right]^{1/2}$$

$$\hbar\omega_\gamma = \left[\frac{\tfrac{1}{2}A^3}{B - \tfrac{3}{2}A^3/(\hbar\omega_\beta)^2}\right]^{1/2} \qquad (2)$$

where, $\hbar\omega_\beta$ and $\hbar\omega_\gamma$ are the head energies of $\beta$- and $\gamma$- vibrations, respectively; and $A$ and $B$ are defined by Eq. (1). Sood[2] used molecular spectra theory[4] to expressed the nuclear moment of inertia dependence on the angular momentum as

$$E(J) = A(J)J(J+1) \qquad (3)$$

where

$$A(J) = A\left[\frac{1 + (a+bJ-1)(B/A)J(J+1)}{1 + (a+bJ)(B/A)J(J+1)}\right].$$

The parameter $(B/A)$ was calculated from the observed energy ratio E(4)/E(2), and the experimental energy of the $2^+$ state was then used to determine $A$, while $a$ and $b$ were adjusted from nucleus to another to give the best fit[2].

---

[*] Physics Department, Al-Fateh University, Tripoli, LIBYA
[†] B-1/1051, Vasant Kunj, New Delhi 110070.

Eq. (3) seems to work well in describing the energies of the ground state bands in rare earth region and the actinides $^{232}$Th and $^{232-238}$U. On the other hand, it does not assume the effect of $\beta$- and $\gamma$- vibrational bands, which are experimentally observed[3] in some nuclei. This fact was encouraging to deduce a new more systematic model by incorporating the Softness parameter[5]

$$\sigma_n = \frac{1}{n!}\frac{1}{\mathcal{J}_0}\frac{\partial^n \mathcal{J}(J)}{\partial J^n}\bigg|_{J=0} \qquad (4)$$

where $\mathcal{J}_0$ is the unperturbed nuclear moment of inertia[6]. Leaving aside few "very rigid nuclei", the inclusion of the $\beta$- and $\gamma$- vibrations along with the variable moment of inertia, has led to the form[7]

$$E(J) = A(J)J(J+1) - B(J)J^2(J+1)^2 \qquad (5)$$

here the rotational parameter $A(J) = \dfrac{\hbar^2}{2\mathcal{J}(J)}$ is connected with to the nuclear variable moment of inertia $\mathcal{J}$, and $B(J) = \dfrac{4A(J)^3}{(\hbar\omega_\gamma)^2} + \dfrac{12A(J)^3}{(\hbar\omega_\beta)^2}$ expresses the correction due to $\beta$- and $\gamma$- vibrations dependence on the nuclear moment of inertia.

**Formalism**
For readability purpose we rewrite Eq. (5) of the form

$$E(J) = A_0 \frac{1}{1+\sigma_1 J} J(J+1) - B_0 \frac{(1-2\sigma_1 J)}{1+\sigma_1 J} J^2(J+1)^2 \qquad (6)$$

where $A_0 = \dfrac{\hbar^2}{2\mathcal{J}_0}$, $\sigma_1$ is defined by Eq. (4), and $B_0$ is related to $A_0$ by Eq. (2).

This non-linear Eq. (6) contains three free parameters $A_0$, $\sigma_1$, and $B_0$ which need to be solved by fit to the experimental data. The equation can be applied to calculate the energy levels of the ground state band of deformed $e$-$e$ nuclei in two major cases:

a) **Three parameters expression**
When the energies of $\beta$- and $\gamma$- vibrations are not experimentally available, then they can be treated as unknown, contained in $c$, and Eq. (6) takes the form

$$E(J) = \frac{A_0}{1+\sigma_1 J} J(J+1) - c A_0^3 \frac{(1-2\sigma_1 J)}{1+\sigma_1 J} J^2(J+1)^2. \qquad (7)$$

The three free parameters $A_0$ and $\sigma_1$, and $c$ can be determined by fitting the first three energy levels $2^+$, $4^+$, and $6^+$ with experimental data.

b) **Two parameters expression**
In some nuclei where the head energies of $\beta$- and $\gamma$- vibrations are experimentally available, then in this case, Eq. (6) can be rewritten as

$$E(J) = \frac{A_0}{1+\sigma_1 J} J(J+1) - \left[\frac{4}{(\hbar\omega_\gamma)^2} + \frac{12}{(\hbar\omega_\beta)^2}\right] A_0^3 \frac{(1-2\sigma_1 J)}{1+\sigma_1 J} J^2(J+1)^2 . \quad (8)$$

This equation contains two free parameters $A_0$ and $\sigma_1$ which can be obtained by fitting only the first two energy levels $2^+$ and $4^+$ with experimental data.

**Results**

The results of our three parametric calculations Eq. (7), compared with experimental data and other existing models are presented in Table 1. The corresponding model parameters $A_0$, $\sigma_1$, and $c$ determined by fitting are given in Table 2.

Keywords:    EXP = Experimental data[8]
             VMI = Variable Moment of Inertia model[9]
             VAVM = Variable Anharmonic Vibrator Model[10], [11]
             GVMI = Generalized VMI[10], [11]
             PM = Present Model (our current work).

Table 1.  Comparison of our results with Experiment and other models in [MeV].

|        | $E(2)$  | $E(4)$   | $E(6)$   | $E(8)$   | $E(10)$ | $E(12)$ | $E(14)$ | $E(16)$ | $E(18)$ |
|--------|---------|----------|----------|----------|---------|---------|---------|---------|---------|
| Ce-150 |         |          |          |          |         |         |         |         |         |
| EXP    | 0.0978  | 0.308    | 0.6087   | 0.9851   |         |         |         |         |         |
| VMI    | ------  | ------   | 0.607    | 0.977    | 1.406   | 1.886   | 2.411   | 2.976   | 3.577   |
| VAVM   | ------  | ------   | ------   | 0.979    | 1.406   | 1.88    | 2.394   | 2.944   | 3.527   |
| GVMI   | ------  | ------   | ------   | 0.983    | 1.419   | 1.907   | 2.422   | 3.02    | 3.634   |
| PM     | ------  | ------   | ------   | 0.9809   | 1.411   | 1.894   | 2.434   | 3.047   | 3.761   |
|        |         |          |          |          |         |         |         |         |         |
| Nd-154 |         |          |          |          |         |         |         |         |         |
| EXP    | 0.0729  | 0.2352   | 0.4789   | 0.807    |         |         |         |         |         |
| VMI    | ------  | ------   | 0.475    | 0.78     | 1.141   | 1.551   | 2.005   | 2.498   | 3.026   |
| VAVM   | ------  | ------   | ------   | 0.792    | 1.166   | 1.591   | 2.062   | 2.574   | 3.122   |
| GVM    | ------  | ------   | ------   | 0.795    | 1.175   | 1.611   | 2.099   | 2.634   | 3.21    |
| PM     | ------  | ------   | ------   | 0.796    | 1.1809  | 1.626   | 2.128   | 2.683   | 3.286   |
|        |         |          |          |          |         |         |         |         |         |
| Sm-154 |         |          |          |          |         |         |         |         |         |
| EXP    | 0.08198 | 0.2667   | 0.5443   | 0.9031   | 1.3333  | 1.8262  |         |         |         |
| VMI    | ------- | ------   | 0.542    | 0.897    | 1.321   | 1.805   | 2.343   | 2.929   | 3.561   |
| VAVM   | ------- | ------   | ------   | 0.901    | 1.327   | 1.81    | 2.346   | 2.927   | 3.549   |
| GVMI   | ------- | ------   | ------   | 0.904    | 1.336   | 1.833   | 2.387   | 2.994   | 3.649   |
| PM     | ------- | ------   | ------   | 0.9032   | 1.3312  | 1.8152  | 2.342   | 2.901   | 3.481   |
|        |         |          |          |          |         |         |         |         |         |
| Sm-156 |         |          |          |          |         |         |         |         |         |
| EXP    | 0.076   | (0.2501) | (0.5179) | (0.878)  |         |         |         |         |         |
| VMI    | -----   | -------  | 0.516    | 0.865    | 1.29    | 1.783   | 2.239   | 2.952   | 3.617   |
| VAVM   | -----   | -------  | -------  | 0.872    | 1.304   | 1.808   | 2.375   | 3       | 3.668   |
| GVMI   | -----   | -------  | -------  | 0.873    | 1.311   | 1.823   | 2.406   | 3.053   | 3.76    |
| PM     | -----   | -------  | -------  | 0.874    | 1.314   | 1.832   | 2.423   | 3.08    | 3.798   |
|        |         |          |          |          |         |         |         |         |         |
| Sm-158 |         |          |          |          |         |         |         |         |         |
| EXP    | 0.0728  | 0.2403   | 0.4985   | (0.8445) |         |         |         |         |         |
| VMI    | ------  | ------   | 0.497    | 0.838    | 1.255   | 1.742   | 2.294   | 2.905   | 3.572   |
| VAVM   | ------  | ------   | ------   | 0.841    | 1.259   | 1.747   | 2.298   | 2.905   | 3.566   |
| GVMI   | ------  | ------   | ------   | 0.842    | 1.265   | 1.762   | 2.328   | 2.957   | 3.645   |
| PM     | ------  | ------   | ------   | 0.842    | 1.265   | 1.759   | 2.317   | 2.93    | 3.588   |

|        | $E(2)$  | $E(4)$  | $E(6)$  | $E(8)$   | $E(10)$ | $E(12)$ | $E(14)$ | $E(16)$ | $E(18)$ |
|--------|---------|---------|---------|----------|---------|---------|---------|---------|---------|
| Gd-160 |         |         |         |          |         |         |         |         |         |
| EXP    | 0.07526 | 0.2482  | 0.514   | 0.868    |         |         |         |         |         |
| VMI    | ------  | ------  | 0.513   | 0.863    | 1.291   | 1.791   | 2.355   | 2.979   | 3.659   |
| VAVM   | ------  | ------  | -----   | 0.86     | 1.29    | 1.79    | 2.35    | 2.96    | 3.62    |
| GVMI   | ------  | ------  | -----   | 0.87     | 1.3     | 1.8     | 2.38    | 3.01    | 3.71    |
| PM     | ------  | ------  | -----   | 0.866    | 1.296   | 1.795   | 2.3519  | 2.954   | 3.589   |
|        |         |         |         |          |         |         |         |         |         |
| Dy-158 |         |         |         |          |         |         |         |         |         |
| EXP    | 0.09894 | 0.31726 | 0.63787 | 1.0441   | 1.5199  | 2.0492  | 2.6126  | 3.1907  | 3.7817  |
| VMI    | ------  | ------  | 0.6355  | 1.0369   | 1.5088  | 2.0417  | 2.6286  | 3.2638  | 3.943   |
| VAVM   | ------  | ------  | ------  | 1.0414   | 1.5135  | 2.0435  | 2.6237  | 3.2485  | 3.9133  |
| GVMI   | ------  | ------  | ------  | 1.045    | 1.5262  | 2.0719  | 2.6749  | 3.3293  | 4.0304  |
| PM     | ------  | ------  | ------  | 1.043    | 1.5162  | 2.0431  | 2.6132  | 3.221   | 3.8663  |
|        |         |         |         |          |         |         |         |         |         |
| Dy-160 |         |         |         |          |         |         |         |         |         |
| EXP    | 0.08679 | 0.28382 | 0.5812  | 0.9672   | 1.4286  | 1.9514  | 2.5152  | 3.0919  | 3.6724  |
| VMI    | ------  | ------  | 0.5807  | 0.9661   | 1.4298  | 1.963   | 2.5587  | 3.211   | 3.915   |
| VAVM   | ------  | ------  | ------  | 0.965    | 1.424   | 1.947   | 2.527   | 3.157   | 3.833   |
| GVMI   | ------  | ------  | ------  | 0.968    | 1.434   | 1.97    | 2.571   | 3.228   | 3.939   |
| PM     | ------  | ------  | ------  | 0.9657   | 1.4213  | 1.9295  | 2.4701  | 3.0211  | 3.5596  |
|        |         |         |         |          |         |         |         |         |         |
| Dy-164 |         |         |         |          |         |         |         |         |         |
| EXP    | 0.07339 | 0.2423  | 0.50132 | 0.84367  | 1.25876 |         |         |         |         |
| VMI    | ------  | ------  | 0.5017  | 0.8453   | 1.2665  | 1.7588  | 2.3164  | 2.9343  | 3.608   |
| VAVM   | ------  | ------  | ------  | 0.8418   | 1.255   | 1.733   | 2.2688  | 2.8569  | 3.4925  |
| GVMI   | ------  | ------  | ------  | 0.8435   | 1.262   | 1.7502  | 2.3022  | 2.9128  | 3.5775  |
| PM     | ------  | ------  | ------  | 0.842    | 1.2532  | 1.7207  | 2.2227  | 2.753   | 3.2751  |
|        |         |         |         |          |         |         |         |         |         |
| Er-162 |         |         |         |          |         |         |         |         |         |
| EXP    | 0.10208 | 0.32963 | 0.66676 | 1.0968   | 1.6028  | 2.1651  | 2.7457  | 3.2923  | 3.8465  |
| VMI    | ------  | ------  | 0.665   | 1.092    | 1.5969  | 2.1702  | 2.8042  |         |         |
| VAVM   | ------  | ------  | ------  | 1.0944   | 1.5975  | 2.1648  | 2.788   |         |         |
| GVMI   | ------  | ------  | ------  | 1.0979   | 1.6104  | 2.1941  | 2.8411  |         |         |
| PM     | ------  | ------  | ------  | 1.0953   | 1.5965  | 2.1518  | 2.7442  | 3.3592  | 3.9857* |
|        |         |         |         |          |         |         |         |         |         |
| Yb-166 |         |         |         |          |         |         |         |         |         |
| EXP    | 0.10238 | 0.3305  | 0.668   | 1.09829  | 1.6059  | 2.1757  | 2.7795  | 3.274   | 3.7831  |
| VMI    | ------  | ------  | 0.6666  | 1.094    | 1.5998  | 2.1738  | 2.8083  |         |         |
| VAVM   | ------  | ------  | ------  | 1.0955   | 1.5978  | 2.1638  | 2.785   |         |         |
| GVMI   | ------  | ------  | ------  | 1.0991   | 1.6109  | 2.1933  | 2.8384  |         |         |
| PM     | ------  | ------  | ------  | 1.096    | 1.5946  | 2.1442  | 2.7265  | 3.3257  | 3.9293* |
|        |         |         |         |          |         |         |         |         |         |
| Yb-168 |         |         |         |          |         |         |         |         |         |
| EXP    | 0.08773 | 0.28655 | 0.5853  | 0.97006  | 1.424   | (1.936) | (2.489) | (3.073) | (3.687) |
| VMI    | ------  | ------  | 0.5854  | 0.9726   | 1.4375  | 1.9713  | 2.5668  | 3.2183  | 3.9209  |
| VAVM   | ------  | ------  | ------  | 0.9692   | 1.4255  | 1.9442  | 2.5175  | 3.1394  | 3.8053  |
| GVMI   | ------  | ------  | ------  | 0.972    | 1.4361  | 1.9688  | 2.5628  | 3.2124  | 3.9182  |
| PM     | ------  | ------  | ------  | 0.9686   | 1.4175  | 1.9097  | 2.4205  | 2.9226  | 3.3874  |
|        |         |         |         |          |         |         |         |         |         |
| Yb-170 |         |         |         |          |         |         |         |         |         |
| EXP    | 0.08426 | 0.27745 | 0.5736  | 0.9636   | 1.4379  | 1.9837  | 2.5808  | 3.1962  | 3.8081  |
| VMI    | ------  | ------  | 0.5723  | 0.9603   | 1.4329  | 1.9822  | 2.6013  | 3.2843  |         |
| VAVM   | ------  | ------  | ------  | 0.961    | 1.431   | 1.974   | 2.582   | 3.248   |         |
| GVMI   | ------  | ------  | ------  | 0.963    | 1.439   | 1.994   | 2.621   | 3.314   |         |
| PM     | ------  | ------  | ------  | 0.964    | 1.441   | 1.992   | 2.604   | 3.264   | 3.957*  |

|        | $E(2)$  | $E(4)$  | $E(6)$  | $E(8)$  | $E(10)$ | $E(12)$ | $E(14)$ | $E(16)$ | $E(18)$ |
|--------|---------|---------|---------|---------|---------|---------|---------|---------|---------|
| Yb-176 |         |         |         |         |         |         |         |         |         |
| EXP    | 0.08213 | 0.27169 | 0.5648  | 0.9541  | 1.4312  | 1.9849  | 2.602   | 3.27    | 3.979   |
| VMI    | ------  | ------  | 0.5641  | 0.9534  | 1.4331  | 1.9965  | 2.6376  | 3.3508  | 4.131   |
| VAVM   | ------  | ------  | ------  | 0.955   | 1.434   | 1.9941  | 2.629   | 3.332   | 4.097   |
| GVMI   | ------  | ------  | ------  | 0.956   | 1.44    | 2.01    | 2.661   | 3.386   | 4.182   |
| PM     | ------  | ------  | ------  | 0.956   | 1.438   | 2.004   | 2.643   | 3.345   | 4.099   |
| Hf-174 |         |         |         |         |         |         |         |         |         |
| EXP    | 0.09101 | 0.29745 | 0.60837 | 1.00943 | 1.487   | 2.022   | 2.599   |         |         |
| VMI    | ------  | ------  | 0.6082  | 1.0111  | 1.4954  | 2.052   | 2.6733  | 3.3534  | 4.0871  |
| VAVM   | ------  | ------  | ------  | 1.009   | 1.4863  | 2.0298  | 2.6315  | 3.285   | 3.9852  |
| GVMI   | ------  | ------  | ------  | 1.0118  | 1.4969  | 2.0546  | 2.6775  | 3.3595  | 4.0955  |
| PM     | ------  | ------  | ------  | 1.0089  | 1.4808  | 2.0031  | 2.5522  | 3.1029  | 3.6282  |
| W-174  |         |         |         |         |         |         |         |         |         |
| EXP    | 0.1119  | 0.355   | 0.704   | 1.137   | 1.635   | 2.186   | 2.78    | 3.392   | 3.973   |
| VMI    | ------  | -----   | 0.7     | 1.14    | 1.65    | 2.22    | 2.84    | 3.51    |         |
| VAVM   | ------  | -----   | -----   | 1.14    | 1.63    | 2.19    | 2.79    | 3.43    |         |
| GVMI   | ------  | -----   | -----   | 1.14    | 1.65    | 2.22    | 2.84    | 3.51    |         |
| PM     | ------  | -----   | -----   | 1.13    | 1.61    | 2.14    | 2.69    | 3.27    | 3.9*    |
| W-178  |         |         |         |         |         |         |         |         |         |
| EXP    | 0.1061  | 0.3431  | 0.6947  | 1.1423  | 1.6661  | 2.2452  | 2.8593  | 3.4891  | 4.1016  |
| VMI    | ------  | ------  | 0.693   | 1.14    | 1.669   | 2.27    | 2.936   | 3.659   |         |
| VAVM   | ------  | ------  | ------  | 1.141   | 1.667   | 2.26    | 2.912   | 3.616   |         |
| GVMI   | ------  | ------  | ------  | 1.145   | 1.68    | 2.29    | 2.967   | 3.704   |         |
| PM     | ------  | ------  | ------  | 1.141   | 1.663   | 2.24    | 2.851   | 3.479   | 4.108*  |
| W-180  |         |         |         |         |         |         |         |         |         |
| EXP    | 0.10357 | 0.33755 | 0.68845 | 1.13847 | 1.66418 | 2.2351  | 2.825   | 3.416   | 4.021   |
| VMI    | ------  | ------  | 0.6879  | 1.1401  | 1.6812  | 2.301   | 2.9911  |         |         |
| VAVM   | ------  | ------  | ------  | 1.1388  | 1.6734  | 2.2806  | 2.9513  |         |         |
| GVMI   | ------  | ------  | ------  | 1.142   | 1.6855  | 2.3088  | 3.0033  |         |         |
| PM     | ------  | ------  | ------  | 1.1386  | 1.6672  | 2.2508  | 2.8639  | 3.48    | 4.072*  |
| W-182  |         |         |         |         |         |         |         |         |         |
| EXP    | 0.10011 | 0.32942 | 0.6805  | 1.1445  | 1.712   | 2.237   | (3.113) |         |         |
| VMI    | ------  | ------  | 0.6792  | 1.139   | 1.6985  | 2.3481  | 3.0797  | 3.8863  | 4.7619  |
| VAVM   | ------  | ------  | ------  | 1.1416  | 1.7008  | 2.3472  | 3.0716  | 3.8662  | 4.7248  |
| GVMI   | ------  | ------  | ------  | 1.144   | 1.7104  | 2.3708  | 3.117   | 3.9422  | 4.8404  |
| PM     | ------  | ------  | ------  | 1.1435  | 1.7066  | 2.356   | 3.0764  | 3.851   | 4.6621  |
| W-184  |         |         |         |         |         |         |         |         |         |
| EXP    | 0.11121 | 0.36406 | 0.74831 | 1.252   | 1.851   |         |         |         |         |
| VMI    | ------  | ------  | 0.7458  | 1.2424  | 1.8408  | 2.5299  | 3.3006  | 4.1455  | 5.0581  |
| VAVM   | ------  | ------  | ------  | 1.2492  | 1.8522  | 2.5448  | 3.317   | 4.1604  | 5.0684  |
| GVMI   | ------  | ------  | ------  | 1.2519  | 1.8629  | 2.571   | 3.3669  | 4.2432  | 5.1932  |
| PM     | ------  | ------  | ------  | 1.2516  | 1.86    | 2.2585  | 3.3313  | 4.1625  | 5.0362  |
| W-186  |         |         |         |         |         |         |         |         |         |
| EXP    | 0.1223  | 0.39647 | 0.80847 | 1.348   | 2.002   |         |         |         |         |
| VMI    | ------  | ------  | 0.8032  | 1.3237  | 1.9424  | 2.6471  | 3.4283  | 4.2784  | 5.1913  |
| VAVM   | ------  | ------  | ------  | 1.3391  | 1.9715  | 2.6919  | 3.4895  | 4.356   | 5.2846  |
| GVMI   | ------  | ------  | ------  | 1.3427  | 1.9854  | 2.7245  | 3.5504  | 4.4548  | 5.4312  |
| PM     | ------  | ------  | ------  | 1.3442  | 1.9902  | 2.7342  | 3.5656  | 4.4758  | 5.4586  |

|        | $E(2)$  | $E(4)$  | $E(6)$  | $E(8)$ | $E(10)$ | $E(12)$ | $E(14)$ | $E(16)$ | $E(18)$ |
|--------|---------|---------|---------|--------|---------|---------|---------|---------|---------|
| Pu-238 |         |         |         |        |         |         |         |         |         |
| EXP    | 0.04408 | 0.14598 | 0.3034  | 0.514  |         |         |         |         |         |
| VMI    | ------- | ------- | 0.3033  | 0.5134 | 0.7727  | 1.0781  | 1.4263  | 1.8143  | 2.2396  |
| VAVM   | ------- | ------- | ------- | 0.5132 | 0.771   | 1.0726  | 1.4142  | 1.7925  | 2.2046  |
| GVMI   | ------- | ------- | ------- | 0.5141 | 0.7744  | 1.0812  | 1.4315  | 1.8223  | 2.2511  |
| PM     | ------- | ------- | ------- | 0.5127 | 0.7689  | 1.0659  | 1.3959  | 1.7501  | 2.1182  |
| Pu-240 |         |         |         |        |         |         |         |         |         |
| EXP    | 0.04283 | 0.14169 | 0.29431 | 0.4976 | 0.7514  |         |         |         |         |
| VMI    | ------- | ------- | 0.2941  | 0.497  | 0.747   | 1.0406  | 1.3746  | 1.7461  | 2.1526  |
| VAVM   | ------- | ------- | ------- | 0.4967 | 0.7446  | 1.0338  | 1.3604  | 1.7212  | 2.1132  |
| GVMI   | ------- | ------- | ------- | 0.4977 | 0.7484  | 1.0432  | 1.3789  | 1.7525  | 2.1616  |
| PM     | ------- | ------- | ------- | 0.4973 | 0.7463  | 1.0358  | 1.3593  | 1.7095  | 2.0779  |
| Pu-242 |         |         |         |        |         |         |         |         |         |
| EXP    | 0.04454 | 0.1472  | 0.3059  | 0.5176 | 0.7787  | 1.0867  |         |         |         |
| VMI    | ------- | ------  | 0.305   | 0.515  | 0.772   | 1.074   | 1.417   | 1.797   | 2.212   |
| VAVM   | ------- | ------  | ------  | 0.517  | 0.776   | 1.079   | 1.422   | 1.802   | 2.216   |
| GVMI   | ------- | ------  | ------  | 0.518  | 0.779   | 1.088   | 1.44    | 1.833   | 2.263   |
| PM     | ------- | ------  | ------  | 0.517  | 0.78    | 1.088   | 1.439   | 1.828   | 2.250   |

* Not reached by other models.

Table 2. Fitting parameters of Eq. (7).

| Nucleus | $\sigma_1 \times 10^{-3}$ | $A_0$ KeV | $c \times 10^{-6}$ KeV$^{-2}$ |
|---------|---------------------------|-----------|-------------------------------|
| Ce-150  | 21.6541                   | 17.14078  | 4.89                          |
| Nd-154  | 15.7048                   | 12.52187  | 0.67                          |
| Sm-154  | 7.9415                    | 13.93509  | 3.48                          |
| Sm-156  | 5.2863                    | 12.81437  | 1.12                          |
| Sm-158  | 2.7119                    | 12.22257  | 2.16                          |
| Gd-160  | 2.4274                    | 12.63628  | 2.67                          |
| Dy-158  | 13.8062                   | 17.04180  | 3.44                          |
| Dy-160  | 4.0815                    | 14.65349  | 3.79                          |
| Dy-164  | 0.2904                    | 12.28586  | 4.24                          |
| Er-162  | 9.9757                    | 17.45010  | 3.18                          |
| Yb-166  | 9.7046                    | 17.49824  | 3.36                          |
| Yb-168  | 3.5835                    | 14.81119  | 4.41                          |
| Yb-170  | 3.0429                    | 14.16695  | 2.26                          |
| Yb-176  | 1.5835                    | 13.75813  | 1.70                          |
| Hf-174  | 3.8362                    | 15.36536  | 3.76                          |
| W-174   | 15.4814                   | 19.40646  | 4.35                          |
| W-178   | 8.9999                    | 18.10533  | 3.02                          |
| W-180   | 5.0998                    | 17.53311  | 3.01                          |
| W-182   | 3.2526                    | 16.84082  | 1.67                          |
| W-184   | 5.8207                    | 18.80577  | 1.41                          |
| W-186   | 12.0535                   | 20.91881  | 0.84                          |
| Pu-238  | -0.0313                   | 7.366827  | 8.59                          |
| Pu-240  | 0.8512                    | 7.168516  | 8.19                          |
| Pu-242  | 2.5625                    | 7.472661  | 4.55                          |

In Table 3, we also present a sample of our calculations in comparison with experimental data and Sood's results[2]. The corresponding parameters $A_0$, $\sigma_1$, and $c$ obtained by fitting are given in Table 4.

Table 3. Comparison of our calculations with the experimental data[2] and Sood's results[2] in [MeV].

|        | $E(2)$  | $E(4)$ | $E(6)$ | $E(8)$ | $E(10)$ | $E(12)$ | $E(14)$ | $E(16)$ | $E(18)$ |
|--------|---------|--------|--------|--------|---------|---------|---------|---------|---------|
| Gd-158 |         |        |        |        |         |         |         |         |         |
| EXP    | 0.07956 | 0.2619 | 0.539  | 0.898  |         |         |         |         |         |
| PM     | 0.07956 | 0.2619 | 0.539  | 0.898  |         |         |         |         |         |
| Sood   | 0.07956 | 0.2617 | 0.539  | 0.901  |         |         |         |         |         |
| Er-166 |         |        |        |        |         |         |         |         |         |
| EXP    | 0.0806  | 0.2649 | 0.545  | 0.910  | 1.334   |         |         |         |         |
| PM     | 0.0806  | 0.2649 | 0.545  | 0.909  | 1.342   |         |         |         |         |
| Sood   | 0.0806  | 0.2647 | 0.544  | 0.908  | 1.344   |         |         |         |         |
| Yb-172 |         |        |        |        |         |         |         |         |         |
| EXP    | 0.0787  | 0.2603 | 0.540  | 0.910  | 1.352   |         |         |         |         |
| PM     | 0.0787  | 0.2603 | 0.540  | 0.910  | 1.361   |         |         |         |         |
| Sood   | 0.0787  | 0.2602 | 0.540  | 0.910  | 1.364   |         |         |         |         |
| Os-184 |         |        |        |        |         |         |         |         |         |
| EXP    | 0.1198  | 0.3836 | 0.774  | 1.274  | 1.871   |         |         |         |         |
| PM     | 0.1198  | 0.3836 | 0.774  | 1.275  | 1.875   |         |         |         |         |
| Sood   | 0.1180  | 0.3834 | 0.777  | 1.278  | 1.867   |         |         |         |         |

Table 4. Fitting parameters of Eq. (7).

| Nucleus | $\sigma_1 \times 10^{-3}$ | $A_0$ KeV | $c \times 10^{-6}$ KeV$^{-2}$ |
|---------|---------------------------|-----------|-------------------------------|
| Gd-158  | -0.14                     | 13.33     | 5.08                          |
| Er-166  | 1.55                      | 13.54     | 4.32                          |
| Yb-172  | 0.12                      | 13.16     | 3.09                          |
| Os-184  | 19.63                     | 20.78     | 0.64                          |

Further application of our model is that, by employing Eq. (8) to calculate the energy levels of ground state band of some deformed *e-e* actinides where the energies of $\beta$- and $\gamma$-vibrations are experimentally available[12]. The results of our calculations are listed in Table 5, and the corresponding values of $A_0$ and $\sigma_1$ determined by fitting are given in Table 6.

Table 5. Comparison of our results with the available experimental data[12] in [MeV].

| | $E(2)$ | $E(4)$ | $E(6)$ | $E(8)$ | $E(10)$ | $E(12)$ | $E(14)$ | $E(16)$ | $E(18)$ |
|---|---|---|---|---|---|---|---|---|---|
| Pu-238 (EXP $E_\beta$ = 0.942 , $E_\gamma$ = 1.029) | | | | | | | | | |
| EXP | 0.044 | 0.146 | 0.303 | 0.513 | 0.773 | 1.079 | 1.427 | 1.816 | 2.241 |
| PM | ------- | ------- | 0.302 | 0.505 | 0.744 | 1.006 | 1.272 | 1.518 | 1.716 |
| Pu-240 (EXP $E_\beta$ = 0.860 , $E_\gamma$ = 0.900) | | | | | | | | | |
| EXP | 0.043 | 0.142 | 0.294 | 0.498 | (0.748) | | | | |
| PM | ------- | ------- | 0.292 | 0.487 | 0.713 | 0.954 | 1.189 | 1.389 | 1.521 |
| U-238 (EXP $E_\beta$ = 0.809 , $E_\gamma$ = 0.927) | | | | | | | | | |
| EXP | 0.043 | 0.143 | 0.296 | 0.497 | 0.741 | 1.024 | 1.341 | 1.688 | 2.063 |
| PM | ------- | ------- | 0.294 | 0.487 | 0.708 | 0.937 | 1.153 | 1.325 | 1.417 |

Table 6. Fitting parameters of Eq. (8)

| Nucleus | $\sigma_1 \times 10^{-3}$ | $A_0$ KeV |
|---|---|---|
| Pu-238 | -3.472 | 7.337 |
| Pu-240 | -4.170 | 7.125 |
| U-238 | -3.157 | 7.257 |

**Conclusion**

The present model Eq. (7) is practically fit to almost all deformed *e-e* nuclei. The results of 24 nuclei where the $\beta$- and $\gamma$-vibrational energies were treated as unknowns are reported in Table 1 and Table 3. A detailed comparison of our calculations with other proposed models, reveals the closest agreement to the experimental energies. At high spin states, as in the case of eg., $^{162}$Er, $^{178}$W, and $^{180}$W nuclei, a surprisingly excellent predictions which had not been satisfactorily described by any other model. Table 1 and Table 3 are direct support to the fact that the $\beta$- and $\gamma$-vibrational energies and the variation of the moment of inertia with spin *J* can not be ignored in evaluating the energies of the ground state rotational bands.

Our model was also applied to some nuclei where the energies of $\beta$- and $\gamma$-vibrations are experimentally available. With the known values of $\hbar\omega_\beta$ and $\hbar\omega_\gamma$, Eq. (7) reduces to two parametric expression Eq. (8). Keeping in view the inherent limitations of Eq. (2), we observe a small deviation at high spin states. Because no much data is available for $\beta$- and $\gamma$-vibrations, it is hoped that experiments may be planned to populate $\beta$- and $\gamma$-vibrational bands to test the validity of our two parametric model and also to verify Eq. (2).